\documentclass[onecolumn,preprintnumbers,amsmath,amssymb]{revtex4}
\usepackage{graphicx}
\usepackage{dcolumn}
\usepackage{bm}
\begin{document}

\begin{center}
{\Large \bf {Wavenumber dependence of the structural relaxation time in the crossover regime of supercooled liquid dynamics} 
\\ \vspace {1cm}}

{ \bf {\large Sarika Maitra Bhattacharyya$^{1}$, Biman Bagchi$^{1,2}$ \\
and \\
 Peter G. Wolynes$^{3}$}}

%\footnote[2] {bbagchi@sscu.iisc.ernet.in}} \\
{$^{1}$ Jawaharlal Nehru Centre for
Advanced Scientific Research
Bangalore 560064,India.}\\
       {$^{2}$ Solid State and Structural Chemistry Unit, 
Indian Institute of Science, Bangalore 560012, India.}\\
%{{(Also at the Jawaharlal Nehru Centre for Advanced Scientific
%Research, Bangalore.)}} \\
{$^{3}$Department of Chemistry and Biochemistry, University of California at San 
Diego, La Jolla, California 92093-0371}\\

\end{center}
\baselineskip=22pt
\begin{center}
{\bf Abstract}
\end{center}
As a liquid is progressively supercooled an intriguing weakening of the wavenumber ($q$) dependence of the structural relaxation time $\tau(q)$ in the large $q$ limit is observed both in 
experiments and simulation studies. Neither the continuous Brownian diffusive dynamics nor the discontinuous 
activated events can alone explain the anomalous wavenumber dependence.
Here we use our recently developed theory that unifies the mode coupling theory (MCT) for continuous dynamics and the random first order transition theory
(RFOT) treatment of activated discontinuous motion as a nucleation (instanton) process, to understand the wavenumber dependence of density relaxation.
The predicted smooth 
change in mechanism of relaxation from diffusive to activated, in the crossover regime, is wavevector dependent
and is eventually responsible for sub-quadratic, almost linear, $q$ dependence of the relaxation time.

%**************************************
\vspace{0.75mm}
%\newpage
\section{Introduction}

Diffusion in an equilibrated supercooled liquid, whatever the mechanism of transport at the microscopic level, should follow Fick's law 
at long length scales and therefore, the relaxation 
time should exhibit a quadratic $q$ dependence at small $q$. However, relaxation at small (or molecular) length scales (large $q$) is sensitive to the detailed mechanism of transport. In deeply supercooled liquid, there is a crossover from continuous (in space) small amplitude Brownian movement of particles 
to rare, nearly discontinuous, larger amplitude hopping motions. This crossover appears to occur over a range of temperature which is limited on one side by the temperature $(T_{L})$ that signifies the onset of barriers for motion in the inherent structures, and ultimately by the glass transition temperature on the other side  when the system falls out of equilibrium \cite{sbp_pnas}. It has been shown that this onset temperature $T_{L}$ \cite{sastry} coincides with the ideal mode coupling theory (MCT) temperature $T_{c}^{o}$ \cite{dave}. In Salol, the range is found to be between 278 K and about 240 K. Above 278K, continuous Brownian diffusion dominates while below 240 K, hopping motion is found to control the mass transport \cite{sbp_pnas}. Long time and long wavelength diffusion is in any event, insensitive to these details of the mass transport or relaxation mechanism found to be present in the crossover region but at intermediate to large $q$, structural relaxation becomes sensitive to the gradually changing mechanism and thus provides a highly useful window to probe the details of the relaxation dynamics. Clearly, at very low temperature (much below MCT $T^{o}_{ c}$ or onset temperature).and in the large $q$ limit, relaxation should become independent of $q$ if mass transport or relaxation is dominated by hopping. 

A characteristic feature of the dynamics in supercooled liquids is a pronounced non-exponential relaxation of the long time decay, commonly known as the 
$\alpha$-relaxation.
This relaxation is usually fitted to the Kohlrausch-William-Watts (KWW) 
function.
\begin{eqnarray}
\phi(q,t)=f(q)e^{-(t/\tau(q))^{\beta(q)}}, \label{KWW}
\end{eqnarray}
\noindent
where $f(q)$,$\tau(q)$, and $\beta(q)$ are the wavenumber dependent 
nonergodicity parameter, the relaxation time and the stretching parameter, respectively.
This functional form exhibits highly non-trivial $q$ dependence. 
When $\phi(q,t)$ denotes the density 
correlator then 
$f(q)$,$\tau(q)$, and $\beta(q)$ oscillates in phase with the static 
structure factor, $S(q)$. Although all the oscillations are in phase with $S(q)$ 
each parameter exhibits different q dependence.  
We may also define another parameter $\Gamma(q)$ which gives the combined effect of 
the $q$ dependence of $\tau(q)$ and $\beta(q)$, $\Gamma(q)=\tau(q)^{-\beta(q)}$. 
The complete $q$ 
dependence 
of the relaxation kernel is given by $\Gamma(q)$. 
If $\Gamma(q)$ has quadratic $q$ dependence, $\phi(q,t)$ will be 
Gaussian. Weaker $q$ dependence of $\Gamma(q)$ leads to non-Gaussian $\phi(q,t)$. 

In the simulation studies of Sciortino and coworkers on water \cite{scior,scior2}
and OTP \cite{scior3}, the variation of $\tau(q)$ with $q$ at different temperatures were studied. 
In their study in the large $q$ limit, as the temperature is lowered, the $q$ dependence of 
$\tau(q)$ shows a crossover from quadratic to linear behaviour,
\begin{eqnarray}
&&\tau(q) \propto q^{-\alpha}.
\end{eqnarray} 
\noindent
At high $T$, $\alpha\simeq 2$ and its value decreases as the temperature is lowered.
These authors also find that $\beta(q)$ varies from 1 at small $q$ to $0.5$ at large $q$. 
In their study, the 
$q$ dependence of $\tau(q)$ is always quadratic in the small $q$ limit \cite{scior}. 
while the dynamics in the large $q$ limit and low temperature is stretched
and non-diffusive \cite{scior,scior2,scior3}. 
Similar subquadratic $q$ dependence at low temperature has been observed in Brownian dynamic simulation 
of hard spheres \cite{sch}. In a different simulation study on polymer melt the authors studied the $q$ dependence of $\tau(q)$ at three different 
temperatures \cite{basch}. In the large $q$ limit, they tried to fit their data to mode coupling theory prediction, $\tau(q)\propto q^{-1/b}$
where $b$ is the von-Schweilders exponent. They find that $b=0.75$ which means that $\tau(q)$ has a weaker than quadratic $q$ dependence. However, if we analyze the results carefully we find that as the temperature is lowered there is further decrease 
in the exponent, i.e the $q$ dependence of $\tau(q)$ becomes even weaker. 
Thus all these studies clearly show a weakening of the $q$ dependence of $\tau(q)$ at low $T$.

According to the idealized mode coupling theory \cite{fuchs,fuchs1} in the large $q$ limit, 

\begin{equation}
\tau(q)^{-\beta(q)}=\Gamma(q)\propto q
\end{equation}
\noindent
IMCT also predicts that $\beta(q)$ varies from 1 at small $q$ and then asymptotically saturates at
 about 0.5 at large $q$, thus
\begin{equation}
lim_{q\rightarrow\infty} \tau(q)\simeq (\Gamma(q))^{-1/\beta(q)}
\propto q^{-1/\beta(q)} \propto q^{-2}
\end{equation}
\noindent
\textit{This would mean that according to IMCT, the dynamics is diffusive even at 
large q.} Although the timescale remains quadratic in $q$, due to 
the stretched dynamics, $(\beta(q)<1)$ $\phi(q,t)$ becomes non-Gaussian. 

The phenomenon of the weakening of the $q$ dependence of $\tau(q)$ as the temperature is lowered is often 
attributed to activated hopping dynamics \cite{arbesim}. However, the jump diffusion model which takes into consideration a 
distribution of jump length, predicts a much weaker $q$ dependence in the large $q$ limit and it also does not account for the non-trivial 
temperature dependence of the exponent $\alpha$ \cite{bookjump}. So neither the continuous diffusive dynamics nor the activated dynamics by itself can explain 
the weakening of the $q$ dependence as observed in simulations \cite{scior,scior2,scior3,sch,basch} 
Both the theories fail to describe the progressive weakening of the $q$ dependence of $\tau(q)$ because in the supercooled liquid the dynamics is neither completely diffusive nor completely activated. As mentioned earlier, the dynamics changes rather smoothly from diffusive to activated and in the crossover region both mechanisms of relaxations are present \cite{sbp_pnas}.

In an earlier study \cite{sbp}, we proposed a unified theoretical approach to combine ideal MCT with a nucleation based random first order transition theory treatment of the activated event. In a subsequent paper,\cite{sbp_pnas} we showed that the unified theory can reproduce many of the observed experimental features, in addition to explaining the crossover behavior mentioned earlier. The theory predicts an interesting interplay between the two mechanisms, in the crossover regime. The earlier study \cite{sbp_pnas} did not address the wavenumber dependence. 
In the present study we investigate the consequences of the theory for the wavenumber dependence of $\tau(q)$ in the large $q$ limit at different temperatures in the crossover regime. 
Here we will study the coherent intermediate scattering function.
In the large $q$ limit apart from the modulation of $\tau(q)$ by 
the structure factor, $S(q)$, the timescale of both the total (coherent) and the self (coherent) intermediate scattering function show similar $q$ dependence. 

The combined theory suggests that an explanation of the wavenumber dependence of $\tau(q)$ must essentially have contributions from both diffusive and activated dynamics as well as an interaction between these two different dynamics which leads to hopping induced diffusion. The earlier study has shown that there is a  gradual change in mechanism of relaxation from diffusive to activated, in the crossover regime. Here we show that change of mechanism of transportation or crossover has a wavenumber dependence. In the intermediate $q$ regime the crossover is sharper and takes place at a higher temperature. This $q$ dependent crossover gives rise to the gradual weakening of the $q$ dependence of the relaxation time. 

\subsection{Theoretical scheme}
The previously developed theoretical scheme that 
combines the hopping (through RFOT theory) and continuous diffusive (described 
using MCT) dynamics can be applied to study the wavenumber dependence of the various dynamical
quantities as described below. 
We also present a pure microscopic derivation of the memory function 

Let us say that an instanton pops up at a position $R$. 
The particles within the sphere of radius $\xi$ around the position $R$
will be displaced by a Lindemann length, $d_{l}$.
Now the master equation describing the change in density $\rho(r)$ due to a 
instanton popping up and due to the continuous diffusive dynamics together can be written as,
\begin{eqnarray}
\delta \rho^{new}({\bf r},t+\delta t)&=& \delta \rho({\bf r},t )+
P_{hop}(r)\times\delta t \frac{1}{V} \int d\textbf{R} \theta ((\textbf{r}-\textbf{R})<\xi)\nonumber\\
&&
\times \Biggl[
\int dt^{\prime}
\int_{{\cal D}(R)} 
d{\bf r}^{\prime} \:\delta \rho({\bf r,}^{\prime}, t^{\prime}) \times G ({\bf r},{\bf r}^{\prime},t-t^{\prime}) 
- \delta \rho ({\bf r},t) \Biggr]\nonumber\\
&&-\delta t \int dt^{\prime} \int dr^{\prime}\delta \rho({\bf r,}^{\prime}, t^{\prime}) K_{MCT}({\bf r},{\bf r}^{\prime},t-t^{\prime}) 
\label{master}
\end{eqnarray}
\noindent
where $\theta(({\bf r} -{\bf R})<\xi)$ provides the effect of the 
instanton felt at the position ${\bf r}$, provided it 
is within the radius $\xi$. $G ({\bf r}, {\bf r}^{\prime}t-t^{\prime})$ is the Greens 
function which determines the effect of instanton 
in moving particles from 
position ${\bf r}$ to a new one ${\bf r}^{\prime}$,
typically a Lindemann length 
away, during time $t$ and 
$t+\delta t$. ${\cal D}(R)$ determines 
the region where the effect of the instanton is felt \cite{sbp}. $P_{hop}(R)$ is the total rate of a hopping in the volume $V$. $K_{MCT}$ is the diffusive kernel which represents 
the diffusive dynamics as described by MCT \cite{sbp_pnas,sbp}.

From Eq.\ref{master} the total scattering function can be written as \cite{sbp},
\begin{equation}
\phi(q,z)=\frac{1}{z+K_{R}(q,z)} \label{fqztot}
\end{equation}
\noindent 
where $K_{R}(q,z)= K_{hop}(q,z)+K_{MCT}(q,z)$ describes the renormalized kernel describing both the activated and the diffusive dynamics. However, in this form we cannot separately analyze the contribution from the diffusive and activated motion. 
Thus to describe
the total intermediate scattering function we use an approximate form by making an approximation,
\begin{equation}
\phi(q,t)\simeq \phi_{MCT}(q,t)\phi^{static}_{hop}(q,t) \label{fqttot}
\end{equation}
\noindent
It has been demonstrated elsewhere that the equation of motion obtained for $\phi(q,t)$ from eq.\ref{fqztot} and 
eq.\ref{fqttot} have very similar form and describe
similar dynamics \cite{sbp}. As mentioned earlier, the advantage of describing $\phi(q,t)$ from eq.\ref{fqttot} 
is that we can now separately analyze the activated and the MCT parts of the total dynamics.   

In describing the activated dynamics we consider that there is a distribution of
the hopping barriers in the system arising from the entropy fluctuation \cite{xiawoly}.
Thus the total contribution from the multiple barrier hopping events is written as,
\begin{eqnarray}{}
\phi^{static}_{hop}(q,t)
&=&\int \phi^{s}_{hop}(q,t) {\cal P}^{static}(\Delta F) d\Delta F \nonumber \\
&=&\int e^{-tK_{hop}(\Delta F)} {\cal P}^{static}(\Delta F) d\Delta F,
\end{eqnarray}
\noindent
where ${\cal P}^{static}(\Delta F)$ is taken to be Gaussian, 
\begin{equation}
{\cal P}^{static}(\Delta F)=
\frac{1}{\sqrt {2{\pi(\delta\Delta F)^{2}}}}e^{-(\Delta F-\Delta F_{o})/2(\delta\Delta F)^{2}}.
\end{equation} 
We call this distribution ``the static barrier height distribution''. 
Here $\phi^{s}_{hop}(t)$ describes the activated hopping dynamics for a single hopping barrier which has been derived in the earlier paper \cite{sbp},
\begin{eqnarray}
\phi_{hop}^{s}(q)=\frac{1}{s+K_{hop}(q,\Delta F)} \label{fqthop}
. 
\end{eqnarray}
\noindent
where $\Delta F$ is the free energy barrier for hopping which determines
the probability of a hop or the waiting time.
\begin{eqnarray}
K_{hop}(q,\Delta F)=\frac{P_{hop}v_{0}}{v_{p}} [1-G(q)] \label{khop}
\end{eqnarray}
\noindent
In the above expression of the hopping kernel, $P_{hop}$ is the average hopping 
rate which is a function of the free energy barrier height, $\Delta F$ 
and is given by $P=\frac {1}{\tau_{0}}exp(-\Delta F/k_{B}T)$ \cite{lubwoly}. 
The 
free energy barrier is calculated from RFOT theory \cite{lubwoly}. 
$v_{0}= \frac{4}{3}\pi \xi^{3}$ 
is the region participating in hopping where the correlation length 
$\xi$ is calculated from RFOT 
theory. $v_{p}$ is the volume of a single particle in the system. $d_{L}$ 
is the Lindemann length. 

Supercooled liquid simulation studies have shown that there is in fact not a single length but rather a distribution of jump lengths 
which can be fitted to an exponential distribution \cite{jumplength}. The most probable jump length has been found 
to be of the order of Lindamann length with the value decreasing as the temperature is lowered. In this work we 
we consider the distribution of jump length, $f_{o}(l)$ to be given by \cite{bookjump},
\begin{equation}
f_{o}(l)=l l_{o}^{-2}exp(-l/l_{o})
\end{equation}
\noindent
then the Greens function can be written as  ,
\begin{eqnarray}
G(q)=\frac{1}{1+q^{2}l_{o}^{2}}
\end{eqnarray}
\noindent
$l_{o}$ is the most probable jump length which is found to be close to the 
Lindemann length.
The hopping kernel can be written as 

\begin{eqnarray}
K_{hop}(q,\Delta F)=\frac{P_{hop}v_{0}}{v_{p}}\left(\frac{q^{2}l_{o}^{2}}{1+q^{2}l_{o}^{2}} \right) \label{khopjump}
\end{eqnarray}
\noindent

We now write the equation of motion for the MCT part of the intermediate  
scattering function, $\phi_{MCT}(q,t)$ which is now self consistently calculated 
with $\phi(q,t)$, 
\begin{eqnarray}
\ddot{\bf\phi}_{MCT}(q,t) 
+\gamma_{q} \dot{\bf \phi}_{MCT}(q,t) 
&+& \Omega_{q}^{2}
{\bf\phi}_{MCT}(q,t) \nonumber \\ 
&+&\int_{0}^{t} \:dt^{\prime}  \Omega_{q}^{2} 
 {\bf m}_{q}^{2}(t^{\prime}) 
\dot{\bf\phi}_{MCT}(q,t-t^{\prime}) = 0 \label{fqtmct}
\end{eqnarray}
\noindent 
where $m_{q}$ is the function of M variables $\phi(q,t)$, (q=1,2,...M) and is
 given by,
\begin{eqnarray}
{\bf m}_{q}(t)&=&{\cal F}_{q}({\bf \phi}(t))\nonumber\\
&=&\sum_{{\vec k}+{\vec p}= {\vec q}}V({\vec q};{\vec k},{\vec p})
{\bf\phi}(k,t) {\bf\phi}(p,t) \nonumber\\
&=&\sum_{{\vec k}+{\vec p}= {\vec q}}[\rho S(q)S(k)S(p)\{{\vec q}[
{\vec k} c(k) +{\vec p} c(p)]\}^{2}/2q^{4}]{\bf\phi}(k,t) {\bf\phi}(p,t) \label{memory}
\end{eqnarray}
\noindent
${\cal F}_{q}$ is the mode coupling functional. The vertices V are equilibrium
quantities \cite{fuchsPRE}. The double summation in Eq.\ref{memory} is approximated by a Riemann sum \cite{fuchsPRE}, where $M=100$ and the cutoff wavevector $q^{cutoff}a=40.0$.

\subsection{Results}
We first discuss the choice of parameters. Following our earlier work, we have taken the thermodynamic conditions appropriate for Salol, although the calculations are quite general.
The system at high temperature is chosen to be just above the idealized MCT transition having $T=280K$, $\rho^{\star}=0.99$ and the lowest temperature $T=229K$, $\rho^{\star}=1.27$.
The last value is suggested by the simulation studies \cite{arnab_sb} of deeply supercooled liquids with model Lennard-Jones potential, which shows that at $\rho^{\star}=1.27$, $T^{\star}=0.5$ the system is highly viscous.
In van der Waals systems the large change in density drives the caging which can in principal arise not only due to change in density but due to transient bond formation which happens in many systems like water. Since we do not explicitly take into consideration the other sources of caging, so in a mean-field way, to mimic strong caging, the density is increased by a large amount. The density between $T=280K$ and $T=229K$ is calibrated in a linear fashion. With these values, we calculate the structure factor using the Percus-Yevik approximation with correction. The structure factor is used in the calculation of the microscopic MCT vertex. The parameters required to calculate the activated dynamics are discussed extensively in Reference \cite{sbp_pnas} 

With the above mentioned parameters $\phi(q,t)$ is calculated at different wavenumbers, $q$, and over the full time domain $t$. 
Numerical calculations were carried out using the scheme given in Ref. \cite{hofac}. 
The longtime part of $\phi(q,t)$ at each wavenumber is fitted to the KWW function (eq.\ref{KWW}), where all the KWW 
parameters, the form factor, $f(q)$, the relaxation time, $\tau(q)$ and the stretching parameter,
$\beta(q)$ are varied to get the fit. Similarly $\phi_{MCT}(q,t)$ and $\phi_{hop}(q,t)$ are also fitted to KWW function
to obtain the corresponding MCT and activated dynamics parameters.

%In figure 1 we plot the relaxation time $\tau_{q}$ as obtained from the present combined theory at high temperature. In figure 2 we %plot the stretching parameter $\beta_{q}$ as a function $q$. We find that wavenumber dependence of $\tau_{q}$ and $\beta_{q}$ trace %each other in phase and also are in phase with the structure factor $S(q)$. Thus at high temperature, the total dynamics behaves in a %similar way as predicted by the idealized MCT \cite{fuchs}.

We plot $log_{10}\tau(q)$ against $log_{10}(q)$ at four different temperatures (\textbf{figure 1}).
As the temperature is lowered, the wavenumber dependence of the relaxation time gradually becomes weak, similar to that observed in experiments \cite{scior,scior2,scior3,sch,basch}.
In \textbf{figure 2} we also plot $log(\tau(q) q^{2})$ vs. $q$ to show the departure from an oscillatory quadratic wavenumber dependence to a sub-quadratic (almost linear) dependence at lower temperatures.
The relaxation time obtained independently from the MCT and the activated parts are plotted in \textbf{figure 3}. In the large $q$ regime, $\tau_{MCT}(q)$ as observed earlier in the case of idealized MCT \cite{fuchs1}, shows a quadratic q 
dependence whereas $\tau_{hop}(q)$ is almost independent of $q$. The quadratic wavenumber dependence  $(\tau(q)\propto \frac{1}{q^{2}})$  is a signature of the continuous Brownian diffusion and the weak wavenumber dependence $(\tau(q)\propto \frac{1}{q^{ \alpha}}$ is a signature of discontinuous activated hopping. Thus neither MCT nor activated dynamics by themselves would explain the subquadratic, almost linear,  wavenumber dependence of the relaxation time. 

In an earlier study we have shown that in the crossover regime there is no abrupt change in the dynamics from diffusive to activated, the change is rather gradual \cite{sbp_pnas}. As we move closer to the laboratory glass transition temperature, the activated dynamics becomes more and more dominant. 
Here we investigate this crossover in the full wavenumber domain. The function $\Delta \tau^{*}=(\tau_{MCT}(q)-\tau_{total}(q))/\tau_{total}(q)$ is plotted as a function of both temperature and wavenumber in \textbf{figure 4}.
The higher the value of this function the stronger is the effect of hopping.
We find that at high temperature, over the whole wavenumber regime the effect of activated dynamics is very weak.
As the temperature is lowered, the effect of the activated dynamics shows strong wavenumber dependence.
In the intermediate $q$ regime the crossover is sharper and takes place at a comparatively higher temperature. 
In the large $q$ regime the crossover is quite gradual.

The wavenumber dependence when activated dynamics is included is largely a consequence of the wavenumber dependence of the 
hopping induced MCT relaxation time. An asymptotic analysis of the MCT dynamics provides the following expression for the timescale of hopping induced diffusive motion below the MCT transition temperature $T_{c}^{o}$\cite{sbp_cond},
\begin{equation}
\tau_{MCT}^{-1}=\frac{2\tau_{hop}^{-1}}{\lambda f(q)^{2}-1}.
\end{equation}
\noindent
Here $f(q)$ denotes the form factor or the Debye Waller factor (DWF). $\lambda$ denotes the coupling strength of the self coupling term and is a function of $m_{q}(t=0)$ (given by eq. \ref{memory}). The above relation between hopping induced MCT relaxation time was derived at a single wavenumber $q=2 \pi/a$, without considering any coupling between wavenumbers. The relation will be more complex for the present more complete treatment where all the wavenumbers are coupled. However it is easy to see that even in the decoupled limit, the relaxation time becomes longer as the DWF increases. Thus at large $q$ although $\tau_{hop}$ is
independent of $q$, the hopping induced diffusive dynamics has a $q$ dependence and the relaxation time is longer for wavenumbers where the DWF is larger.

In \textbf{figure 5} the total relaxation time, $\tau_{total}$, the hopping induced MCT relaxation time, $\tau_{MCT}$ and the relaxation time of the activated hopping dynamics, $\tau_{hop}$ are plotted at three different temperatures.
At high $T$, for all $q$, as discussed earlier, the dynamics is primarily determined by MCT. However, 
at low $T$ and in the intermediate wavenumber regime the MCT relaxation time becomes very slow (due to caging effect) thus the relaxation primarily occurs via activated dynamics. As observed in \textbf{figure 5}, in the large wavenumber regime, even at low $T$ 
both MCT and activated dynamics
have competing timescales and the relaxation has contributions from both. This wavenumber dependence of the effect of hopping 
and stronger effect of activated dynamics at intermediate wavenumbers would imply that the growth of relaxation time with lowering of $q$ will not be as strong (quadratic) as predicted by the MCT dynamics.

This theory also predicts that even in the supercooled regime there is a limited range of $q$ where non-Gaussian 
behaviour can be observed. This range, which is primarily positioned in the intermediate wavenumber regime, becomes asymmetrically wider as the temperature is lowered. As observed in \textbf{figures 4 and 5}, the effects of hopping dominated regime extends to higher $q$ values with lowering of $T$. Thus the 
non-Gaussian regime also extends to higher values of $q$.
However in the lower wavenumber side even when the hopping dominated regime is reached the non-Gaussian wavenumber dependence will be restricted only above $q*a\simeq 2\pi $. This is because, the activated dynamics itself predicts a Gaussian wavenumber dependence at small $q$.

\subsection{Discussion}
The unusual wavenumber dependence of the relaxation time observed in supercooled liquids has heretofore remained largely unexplained \cite{scior,scior2,scior3,sch,basch}.
This article addresses this problem by extending the bridged theory\cite{sbp_pnas,sbp} to the full wavevector plane. We show that the full wavevector dependent unified 
theory explains the gradual weakening of wavenumber dependence of the relaxation time as the temperature is lowered. The theory predicts that for systems where the caging effect grows strongly with lowering of temperature, the MCT relaxation time in the intermediate wavevector regime slows down faster and the crossover to the activated dynamics dominated regime happens earlier. We have shown that this crossover to the purely activated regime is strongly wavenumber and temperature dependent. The crossover to the activated regime 
happens earlier at intermediate wavenumbers, whereas at the same temperature and at larger wavenumbers, the MCT contribution still remains
substantial. Because of this wavenumber dependent crossover the relaxation time at low $T$ is governed mostly by activated dynamics at intermediate $q$ and both by activated and MCT dynamics at larger $q$. Hence the relaxation time does not grow as strongly with lowering of $q$ as expected from only the diffusive MCT dynamics. The theory also predicts that if the caging effect is not so strong so as to  significantly slow down the MCT dynamics (compared to hopping) we should observe quadratic wavenumber dependence even at lower temperature.  
The analysis also reveals that for systems where the caging effect is stronger, the transition from MCT dominated regime to the activated dynamics dominated regime should be sharper. This would mean that the phenomenological MCT transition temperature, $T_{c}^{fit}$ will appear to be closer to the microscopic MCT transition temperature $T_{c}^{0}$. In other cases where the caging effect is not very strong or does not grow very fast upon cooling the transition takes place over a wider temperature regime and the separation between $T_{c}^{fit}$  and $T_{c}^{0}$ should be wider.

It is now believed that supercooled liquids are characterized by a growing dynamic correlation length, $\xi$. For activated transitions in pure RFOT theory, the growth of this correlation length is given by the configuration entropy, $\xi \propto s_{c}^{-2/3}$ \cite{lubwoly}. At high temperature where this dynamic correlation length is small, relaxation is described by the MCT. As the temperature is lowered, $\xi$ grows, and as discussed here, with the lowering of $T$, activated dynamics also becomes progressively a major contributor to the total relaxation dynamics. It is thus reasonable to ask if there is a correlation between $\xi$ and 
the non-Gaussian wavenumber dependence. The present analysis suggests that while a growing correlation length is essential to describe the timescale of activated dynamics and thus the non-Gaussian dependence is related to $\xi$, there appears to be no one-to-one relation between $\xi$ and the non-Gaussian dependence for the following reason. The non-Gaussian dependence is not described by activated dynamics alone but it is MCT and activated dynamics and their interplay which gives rise to the non-Gaussian wavenumber dependence of the 
relaxation time. Since MCT dynamics is independent of $\xi$ and depends on many other factors like caging, 
the relationship between the correlation length $\xi$ and the non-Gaussian wavenumber dependence may not be universal.
For the system we have studied, it is fair to say that sub-quadratic wavenumber dependence becomes significant only when the correlation length $\xi \ge 2$.

\textbf{Acknowledgment}-SMB thanks JNCASR(INDIA), BB thanks DST (INDIA) and PGW thanks NSF (USA) for funding. SMB thanks Dr. F. Sciortino for discussions.

\newpage

%\begin{figure}
%\includegraphics[width=18cm]{tauq_highT_fig1.pdf}
%\caption{The $\alpha$ relaxation timescale $\tau_{q}$ plotted as a function of
%$q$ at T=280Kand $\rho^{*}=0.99$. The $\tau_{q}$ value oscillates in phase  with S(q).}
%\end{figure}

%\begin{figure}
%\includegraphics[width=18cm]{betaq_highT_fig2.pdf}
%\caption{The stretching parameter $\beta_{q}$ plotted as a function of
%$q$ at T=280Kand $\rho^{*}=0.99$. The $\beta_{q}$ value oscillates in phase with S(q).}
%\end{figure}

\begin{figure}
\includegraphics[width=18cm]{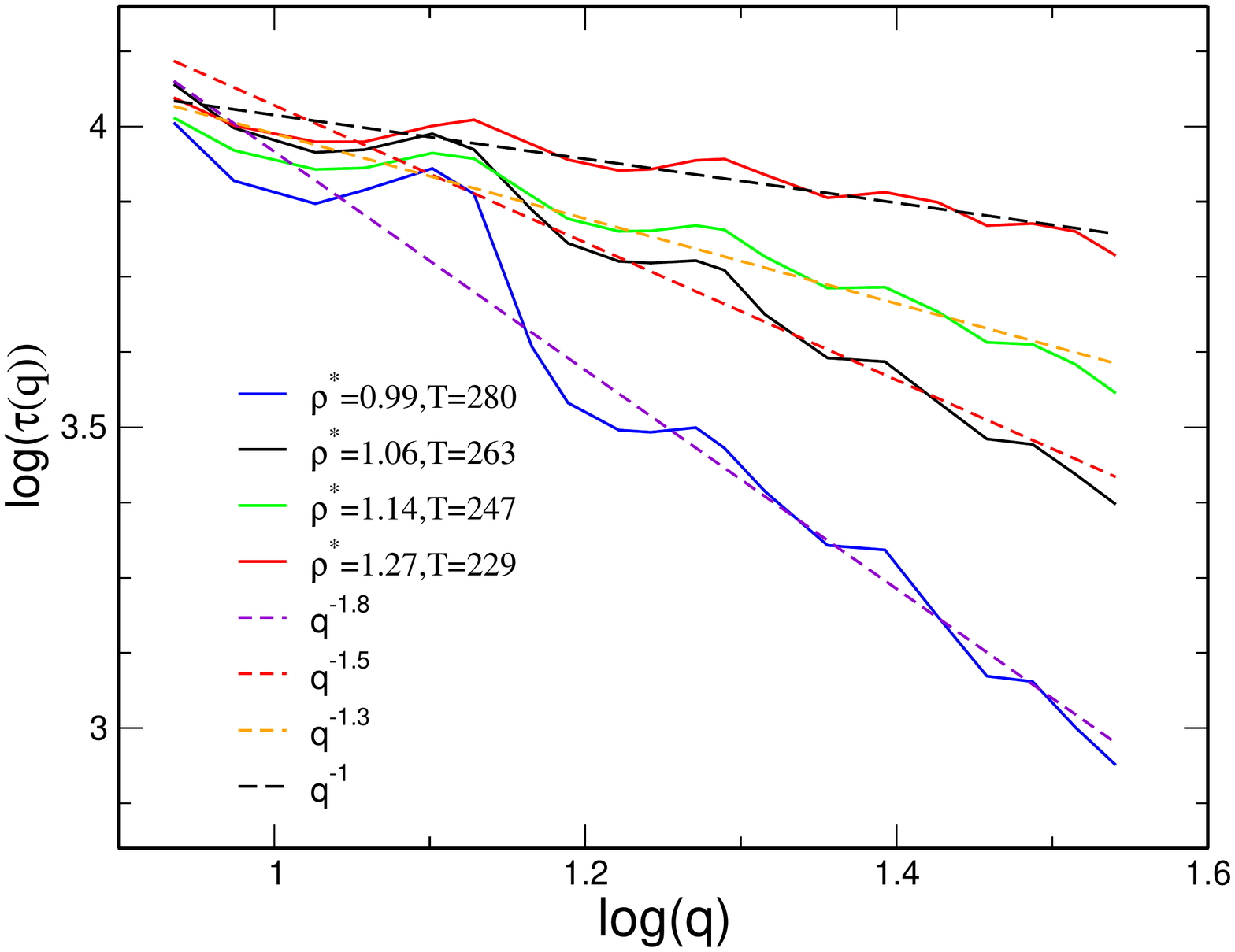}
\caption{The $\alpha$ relaxation timescale $\tau(q)$ plotted as a function of
$q$ at different densities and temperatures. The $\tau(q)$ values are scaled such that at $q=8.6$ they have similar values. $\tau(q)$ shows a weaker $q$ dependence as the temperature is lowered.}
\end{figure}

\begin{figure}
\includegraphics[width=18cm]{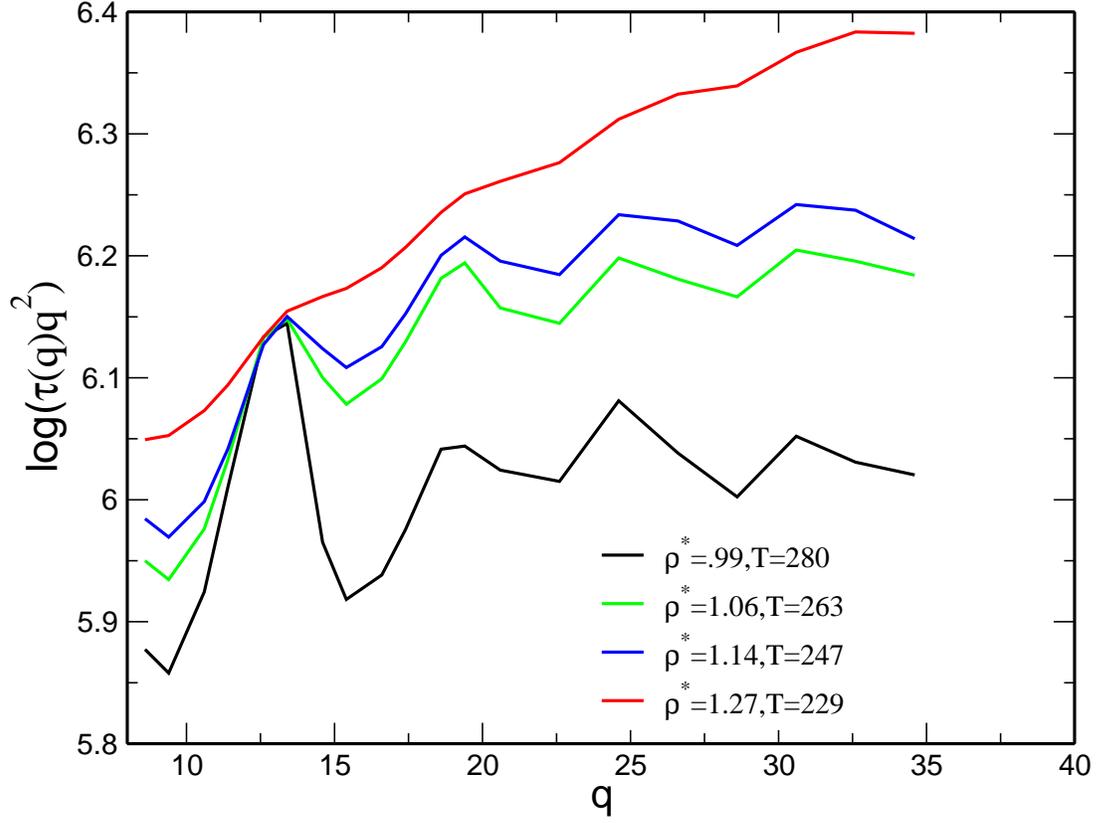}
\caption{ $\tau(q) q^{2}$ plotted as a function of
$q$ at different densities and temperatures. The $\tau(q)q^{2}$ values are scaled such that at $q=12.5$ they have similar values. As the temperature is lowered, $\tau(q)$ shows a crossover from quadratic to sub-quadratic (almost linear) $q$ dependence.} 
\end{figure}

\begin{figure}
\includegraphics[width=18cm]{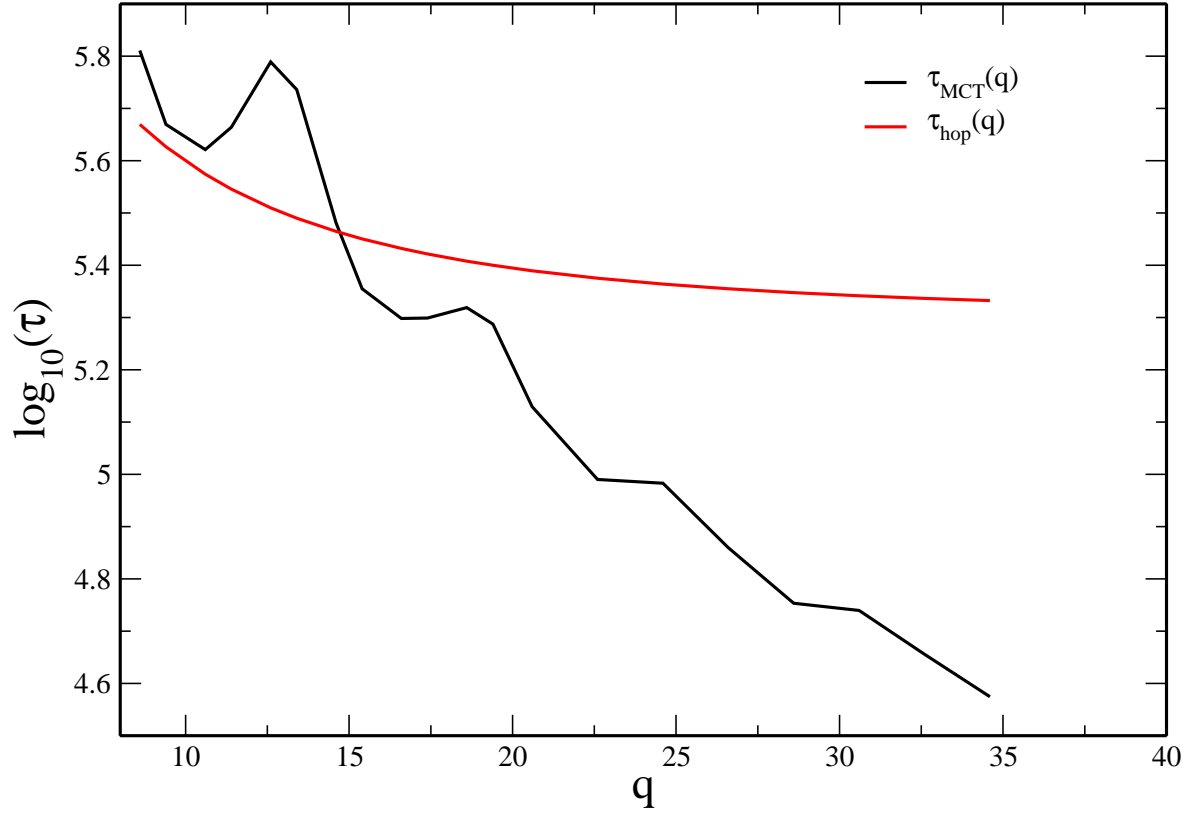}
\caption{The $\alpha$ relaxation timescale $\tau(q)$ for the MCT part, $\tau_{MCT}$ and the activated part, $\tau_{hop}$ are 
plotted as a function of $q$. $\tau_{MCT}$ shows a strong variation with $q$ whereas $\tau_{hop}$ has a much weaker $q$ dependence. The relaxation times are calculated at $T=263 K$ and $\rho=1.16$}
\end{figure}

\begin{figure}
\includegraphics[width=15cm]{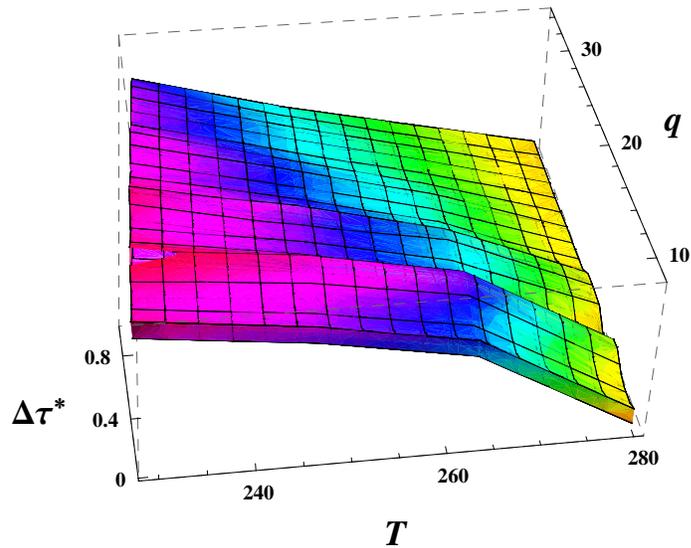}
\caption{Wavenumber dependent crossover to the activated regime.
The function $\Delta \tau {*}=(\tau_{MCT}-\tau_{total})/\tau_{MCT}$ is plotted against both temperature and wavenumber.
The closer the value of the function to unity the more is the effect of activated dynamics. In the intermediate wavenumbers, the crossover to the activated dynamics dominated regime is sharper and takes place at a higher temperature. In the large $q$ regime this crossover takes place over a wide temperature range.}
\end{figure}

\begin{figure}
\includegraphics[width=14cm]{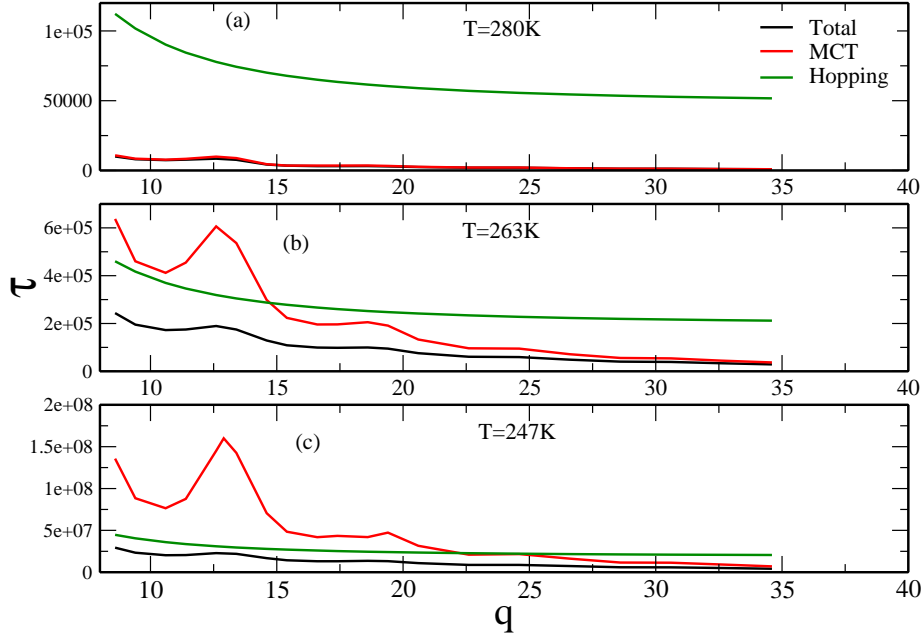}
\caption{(a)The total relaxation time, $\tau_{total}$, the hopping induced MCT relaxation time, $\tau_{MCT}$ and the relaxation time of the activated hopping dynamics, $\tau_{hop}$ are plotted against time at T=280K. The total relaxation is primarily determined by the hopping induced diffusive dynamics.
(b)Same as in (a) but at T=263K. The total relaxation in the intermediate wavenumber is primarily determined by the hopping dynamics whereas in the long $q$ regime the total dynamics follows the relaxation timescale of the hopping induced diffusive dynamics.
(c)Same as in (a) but at T=247K. Here too the total relaxation in the intermediate wavenumber is primarily determined by the hopping dynamics and only in the very large wavevector regime MCT plays a dominant role.}
\end{figure}
\end{document}